# Phase transitions driven by solute concentration, temperature, and pressure in uranium-6wt % niobium alloy

Yanwen Liao[a], Yongfeng Huang[a], Kun Wang[a] *, Wenjun Zhu[b], Wu-Xing Zhou[c], Yi Liao[d], Songlin Yao[b] *

[a] *College of Materials Science and Engineering, State Key Laboratory of Cemented Carbide, Hunan University, Changsha 410082, China*

[b] *National Key Laboratory of Shock Wave and Detonation Physics, Institute of Fluid Physics, Mianyang 621900, China*

[c] *School of Materials Science and Engineering, Hunan University of Science and Technology, Xiangtan, 411201, China*

[d] *School of Mechanical Engineering, Southwest Petroleum University, Chengdu, 610500, China*

**Abstract**

An angular-dependent potential for the U-Nb system is developed based on an existing ADP for U and a new EAM potential for Nb through fitting flexible cross-interaction functions and alloy parameters to experimental and first-principles data, enabling accurate prediction of phase transitions ($\alpha \leftrightarrow \gamma$ driven by solute concentration; $\alpha'' \to \gamma$ under temperature in U-6Nb alloy), elastic properties, defect energetics, and mixed enthalpy. The potential reliably reproduces melting points of U-Nb solid solutions and captures lattice parameter expansion of $\gamma$ U-6Nb. Notably, it correctly predicts Hugoniot relations and equations of state up to ~90 GPa and resolves the $\alpha'' \to \gamma$ transition under static high pressure. Combined with atomic simulations, we reveal a twinning-coupled $\alpha'' \to \gamma$ transition of U-6Nb under high pressures: $\{112\}_\gamma$ twins form via nanosecond-scale twinning precursors generated during the transient adiabatic compressions. The static phase transition pressure is predicted to be 54.5GPa, comparable to 67.2 GPa by first principle calculations. Besides, our result suggests that U-

---

* Corresponding authors. E-mail: kwang_hnu@163.com (K. Wang), yaosl@caep.cn (S. Yao)

6Nb single-crystal would experience a nonlinear elastic relaxation before yielding plastically at 3.1 GPa (shear stress: 0.9 GPa). The results in this work help resolve long-standing discrepancies in understanding the abnormal shear stress relaxation mechanisms under high-pressure shock loading.

## I. INTRODUCTION

Uranium is extensively utilized in weapon systems and nuclear reactors due to its high density and unique nuclear properties. Three crystalline phases of pure uranium have been identified: the orthorhombic $\alpha$-phase at low temperatures, and two high-temperature phases - the orthorhombic $\beta$-phase and the body-centered cubic (BCC) $\gamma$-phase at elevated temperatures [1, 2]. The environmentally stable $\alpha$-phase exhibits poor corrosion resistance and an unfavorable combination of strength and ductility, posing significant challenges for engineering applications. While the $\gamma$-phase demonstrates the most favorable engineering properties, it unfortunately cannot be retained under ambient conditions. To stabilize the $\gamma$-phase at room temperature, the prevailing strategy involves alloying uranium with transition metal elements (e.g., Nb [3], Mo [4], Ti [5], Zr [6]). These elements form infinite solid solutions with $\gamma$-U at high temperatures but exhibit near-zero solubility in low-temperature $\alpha$-U. This limitation can be circumvented through $\gamma$-phase heat treatment of uranium alloys followed by rapid quenching to room temperature. The high cooling rate induces martensitic transformation of the $\gamma$-phase, resulting in a metastable supersaturated solid solution. Particularly noteworthy is the U-Nb system, which demonstrates significantly enhanced corrosion resistance, reduced Young's modulus, and improved toughness compared to pure uranium [7, 8]. The U-Nb alloy system exhibits complex phase behavior with multiple competing stable and metastable phases depending on temperature and Nb content. The high-temperature $\gamma$-phase shows complete solubility in the equilibrium U-Nb phase diagram, undergoing equilibrium segregation reactions near 923 K [9]. As these segregation processes involve sluggish solid-state diffusion, they can be effectively suppressed by moderate cooling rates ($\geq$ 20 K/s) [8]. Industrially, U-Nb alloys are typically processed through $\gamma$-phase solution treatment followed by quenching. The system features a rich metastable phase diagram: At low Nb concentrations (0-3.5 wt.%), direct $\gamma \rightarrow \alpha'$ transformation occurs, where $\alpha'$ shares the orthorhombic structure of $\alpha$-phase with minor lattice parameter differences. Intermediate Nb levels (3.5-6.5 wt.%) induce sequential transformations: $\gamma \rightarrow \gamma^o$ (orthorhombically distorted $\gamma$-phase with $c/a \approx 0.97$) $\rightarrow \alpha''$ (monoclinically distorted $\alpha'$ phase with $\gamma$-angle deviating 1-3° from 90°). Higher Nb content (6.5-9 wt.%) suppresses $\gamma^o \rightarrow \alpha''$ transformation, retaining $\gamma^o$ as the martensitic product. Beyond ~9 wt.% Nb, the $\gamma$-phase becomes fully retainable at ambient conditions. Recent studies have investigated martensitic start ($M_s$) and austenitic start ($A_s$) temperatures in quenched U-Nb alloys with varying Nb content [10]. Notably, both transformation temperatures demonstrate significant cooling/heating rate dependence [11].

Among various Nb-containing alloys, the U-6wt.% Nb alloy (U-6Nb) demonstrates exceptional strength-ductility synergy, corrosion resistance, and shape memory effect, making it particularly valuable for weapon systems and nuclear reactor applications. U-6Nb has been designated as a strategic reserve material [10], attracting substantial research attentions. Industrially, U-6Nb is typically processed through solution treatment followed by quenching: homogenization at ~1053-1273 K followed by rapid water cooling to obtain the metastable $\alpha''$ phase. During quenching, U-6Nb undergoes sequential phase transformations at ~600 K ($\gamma \rightarrow \gamma^o$) and ~390 K ($\gamma^o \rightarrow \alpha''$) [10].

Extensive experimental studies using X-ray/neutron diffraction have been conducted to characterize phase transformation behavior in U-6Nb during quenching and aging. However, such diffraction techniques may fail to capture atomic-scale structural features—a phenomenon previously noted in U-transition metal alloy systems. For instance, the "correlated disorder" observed in U-Mo alloys generates locally distorted structures. The weak diffuse scattering signals from such atomic-scale correlations are easily obscured in conventional diffraction patterns, leading to frequent oversight of these features [12]. The performance stability of U-6Nb in service environments is intrinsically linked to the as-quenched $\alpha''$ phase, which remain incompletely resolved. Among the key features of U-6Nb alloy, martensitic phase transitions due to the potential variations of composition, temperature and pressure are of particular importance. For example, the transitions are hypothesized to suppress elastic behavior under external loading by leveraging volumetric changes during the phase transformation to relieve shear stresses—a mechanism distinct from conventional elastic-plastic responses [13-19]. Notably, the role of twinning observed experimentally remains unclear in the scenarios where the phase transitions predominantly govern the shear stress relaxation. Current understanding on the martensitic transformation mechanisms relies primarily on crystallographic geometric deductions, with limited atomistic-scale evidence available [20, 21]. These knowledge gaps persist due to the inherent experimental challenges posed by its high reactivity and radioactivity. The limitations of experimental methodologies have driven increasing adoption of computational simulation approaches. Molecular dynamics (MD) simulations provide atomic-scale spatial resolution and picosecond-level temporal resolution, making them particularly suitable for investigating diffusionless martensitic transformations. However, the reliability of MD simulations critically depends on the accuracy of interatomic potentials. At present, quantities of interatomic potentials are developed for α-U by different authors in different application fields, for example, the Charge Optimized Many-Body (COMB) potential [22], Embedded Atom Method (EAM) potential [23], Modified Embedded Atom Method (MEAM) potential [24] and machine-learning-based Moment Tensor Potential (MTP) [25] among others. Specially, the EAM potential by Smirnova et al. [23] can well predict the deformation twinning in α-U single crystals under shock compressions, but the pressure threshold for the phase transition from α-U to BCT-U is implausible [26]. Regarding Nb, numerous potentials exist, such as Finnis-Sinclai (FS) [27], EAM [28-30], MEAM [31, 32] and Long-Range Empirical Potential (LREP) [33]. Zhang et al. [34] compare these potentials and find that the FS [27] and EAM [28] potential could reproduce twinning phenomena induced by high pressures and the generalized stacking fault energy curve. Li et al. [30] specially developed a new EAM potential for describing the shock responses of Nb under high pressures. With regards to the U-Nb system, Dai et al. developed a long-range U–Nb potential [33] for describing vacancies and surface properties. Recently, Huang et al. constructed a machine-learning based moment tensor potential [35] to study the $\gamma \rightarrow \gamma^0 \rightarrow \alpha''$ martensitic transformation in U-6Nb under quenching. While the machine learning interatomic potentials could in principle simulate the phase transitions in U-6Nb alloys, practical implementations are currently constrained to systems of ~10,000 atoms due to their high computational costs.

In this study, an angular-dependent potential (ADP) for the U-Nb alloy is carefully developed and thoroughly tested, enabling efficient MD simulations of systems containing tens of millions of atoms without the need for extensive computational resources. Further, the role of phase transitions in governing the mechanical behavior of U-6Nb is systematically investigated through MD simulations using Large-scale Atomic Massively Parallel Simulator [36]. The remainder of this work is structured as follows: In Part II, we detail the development of the U-Nb alloy potential and validate its accuracy by evaluating numerous physical properties critical to shock responses. Part III explores the martensitic phase transitions in U-6Nb alloys induced by compositional fluctuations, temperature, or pressure. Finally, conclusions are summarized in Part IV.

## II. DEVELOPMENT OF U-Nb POTENTIAL

### A. First principle calculations

First principle calculations of energy and structure are performed within the Density Functional Theory (DFT) [37, 38] [39, 40]framework, using the Vienna Ab initio Simulation Package (VASP) [41]. The Projector Augmented Wave (PAW) method [42, 43] and the Perdew-Burke-Ernzerhof (PBE) [44] parameterization within the Generalized-Gradient Approximation (GGA) are used to describe electron-ion interactions and exchange-correlation interactions, respectively. The cutoff energy is set to 600 eV, with an energy tolerance of $10^{-8}$ eV for the electronic self-consistent steps and the Hellmann–Feynman force convergence criterion of -0.001 eV/Å per atom. The K-point grid is sampled using a $\Gamma$-centered mesh with a spacing of 0.1 Å$^{-1}$ in reciprocal space. The valence electron configurations of U($6s^26p^65f^36d7s^2$) and Nb($4s^24p^64d^45s$) are chosen for the two elements. The convergence parameters ensure that the calculated energy fluctuations of atom configurations involved in this work are within 1 meV/atom. Given that Nb is non-magnetic and U exhibits Pauli paramagnetic with very weak magnetism (<0.005 μB/atom) [45, 46], the U-Nb alloys demonstrate similar behavior. Therefore, all calculations are performed without considering magnetic effects.

For solution alloys, lattice distortion due to the substitution of solution atoms plays a key role in mechanical behaviors in response to the external loadings. When constructing the ADP potential for the U-Nb alloy, we choose to fit both energies and pressures of a serial of atom configurations involving γ-phase U-Nb substitutional solid solutions and α"-U$_7$Nb$_1$ under both regular and compressed states. For the γ-phase substitutional solution, all possible random atom configurations are enumerated using a *derivative-structure-enumeration library* [47]. Additionally, a series of γ-phase stacking fault and twin configurations with different solution concentrations are taken into account because of their importance in the mechanical responses of U-Nb alloys under external loadings. During constructing the potential, we temporarily disregard lattice distortion from elemental substitution in the solid solution, focusing only on the effects of elemental differences on system energy and pressure. This is because when using the constructed ADP potential, lattice distortion naturally forms upon relaxation of the solid solution. Therefore, the first-principles calculations on the solid solutions are performed without lattice relaxations. According to our practical experience, using excess mixture energy ($\bar{U}_{exc}$) to describe the alloy effects of the atom configurations under different solution concentrations is a good choice. The excess mixture energy is defined by

$$\bar{U}_{exc} = \bar{U}_{tot} - \sum_{i=1}^{n} \omega_i \bar{U}_{(t_i)}, \tag{1}$$

where $\omega_i$ represents the atomic fraction of the *i*-th element, $\bar{U}_{tot}$ represents the per-atom energy of the solution alloy, and $\bar{U}_{(t_i)}$ represents the per-atom energy of the same alloy configuration but replaced entirely by the *i*-th element. Note that all the energies above are calculated without lattice relaxation.

The DFT calculation data for the chosen U-Nb alloy configurations are shown in TABLE I. Configuration U(γ)-1 represents γ-U phase with eight atoms at the mechanical equilibrium state. Configuration U(γ)-2 represents a uniformly compressed γ-U phase (8 atoms) with a linear compression ratio of 0.98. Subsequently, U(γ)-3 and U(γ)-4 denote the γ-U phase (96 atoms) and its corresponding {110} <111> unstable stacking fault (USF) configuration; U(γ)-5, U(γ)-6, and U(γ)-7 denote the γ-U phase (72 atoms) and its corresponding {112} <111> USF and {112} <111> twinning configurations. When doping Nb in these USF and twinning configurations, overall random doping and doping within or near the slip plane (within a cutoff distance of 6.9 Å) were considered. Given the large size of USF and twinning configurations and our focus on their relative energy differences with the perfect configurations rather than their absolute values, the DFT cutoff energy, electronic self-consistency tolerance, force convergence criterion can be slightly adjusted to 500 eV, $10^{-6}$ eV and -0.01 eV/Å per atom, respectively, with a reciprocal space interval of 0.2 Å$^{-1}$, to improve computational efficiency while maintaining sufficient accuracy. Then, configuration U($\alpha''$)-1 is $\alpha''$-U$_7$Nb$_1$, constructed as described in Refs. [48, 49]. Atomic fractional coordinates are as follows: (0.05, 0.10, 0.75), (0.48, 0.12, 0.75), (0.27, 0.41, 0.25), (0.84, 0.40, 0.25), (0.20, 0.59, 0.75), (0.96, 0.86, 0.25), (0.52, 0.90, 0.25), (0.69, 0.63, 0.75). Configuration U($\alpha''$)-2, U($\alpha''$)-3, U($\alpha''$)-4, U($\alpha''$)-5 and U($\alpha''$)-6 represent $\alpha''$-U$_7$Nb$_1$ subjected to a uniform compression with a linear compression ratio of 0.98, 0.96, 0.94, 0.92, and 0.90, respectively. The atom configurations of $\gamma$-U, its stacking fault and twin configurations, as well as $\alpha''$-U$_7$Nb$_1$, are visualized in FIG. 1 using VESTA [50]. More results about U-Nb alloys with varying solution concentrations will be discussed alongside the later MD simulation results.

TABLE I. The DFT calculation data for U-Nb alloy configurations, where the reference lattice parameters are collected from experiments [51, 52] or first principles calculations reported in literatures [48, 49, 53].

| Structure | System | $P_0$ (GPa) | Lattice Parameter | |
| --- | --- | --- | --- | --- |
| | | | This work | Reference |
| U($\gamma$)-1 | Cubic | 0 | 3.432 Å, 6.864 Å, 6.864 Å | 3.47 Å, 6.94 Å, 6.94 Å [51] |
| | | | | 3.43 Å, 6.86 Å, 6.86 Å [53] |
| U($\gamma$)-2 | Cubic | 9.19 | 3.363 Å, 6.727 Å, 6.727 Å | |
| U($\gamma$)-3 | Cubic | -0.14 | 8.409 Å, 38.839 Å, 5.946 Å | |
| U($\gamma$)-4 | Cubic | 0.08 | 8.409 Å, 38.839 Å, 5.946 Å | |
| U($\gamma$)-5 | Cubic | -0.18 | 4.855 Å, 50.453 Å, 5.946 Å | |
| U($\gamma$)-6 | Cubic | 1.08 | 4.855 Å, 50.453 Å, 5.946 Å | |
| U($\gamma$)-7 | Cubic | -0.05 | 4.855 Å, 50.453 Å, 5.946 Å | |
| U($\alpha''$)-1 | Monoclinic | 0 | 5.968 Å, 5.705 Å, 4.886 Å, 96.5 ° | 5.701 Å, 5.765 Å, 4.997 Å, 91.3 °[52] |
| | | | | 5.698 Å, 5.760 Å, 4.994 Å, 91.3 °[49] |
| | | | | 5.958 Å, 5.699 Å, 4.885 Å, 96.3 °[48] |
| U($\alpha''$)-2 | Monoclinic | 10.25 | 5.849 Å, 5.591 Å, 4.788 Å, 96.5 ° | |
| U($\alpha''$)-3 | Monoclinic | 24.02 | 5.729 Å, 5.477 Å, 4.691 Å, 96.5 ° | |
| U($\alpha''$)-4 | Monoclinic | 42.35 | 5.610 Å, 5.363 Å, 4.593 Å, 96.5 ° | |
| U($\alpha''$)-5 | Monoclinic | 66.68 | 5.491 Å, 5.249 Å, 4.495 Å, 96.5 ° | |
| U($\alpha''$)-6 | Monoclinic | 98.95 | 5.371 Å, 5.135 Å, 4.397 Å, 96.5 ° | |

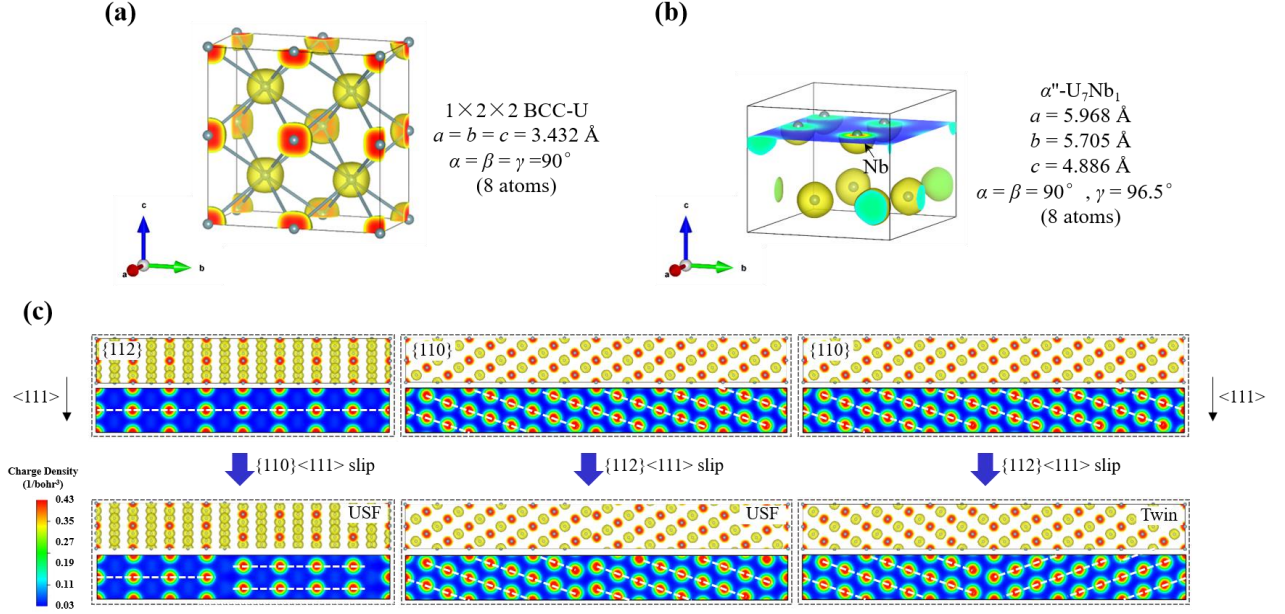

FIG. 1. Atomic arrangements and charge density distributions of (a) $\gamma$-U (isosurface level is 0.30 $a_0^{-3}$), (b) $\alpha''$-U$_7$Nb$_1$ (isosurface level is 0.31 $a_0^{-3}$), and (c) unstable stacking fault and twinning configurations (isosurface level is 0.30 $a_0^{-3}$) used for potential fitting. Specifically, the charge density distribution at the z = 0 height slice on the {112} or {110} plane is also shown in figure (c). $a_0$: Bohr radius.

### B. Angular-dependent potential and parameterization

According to the framework of the ADP firstly proposed by Mishin et al [54], the total energy ($E$) of the simulated system is expressed as the summation of pairwise interaction ($\phi_{(t_i t_j)}$), embedding energy ($F_{(t_i)}$) and the energy modified terms for non-central interactions, relevant to the local dipole vectors ($\mu_i^\alpha$) and the quadrupole tensors ($\lambda_i^{\alpha\beta}$), that is

$$E = \frac{1}{2}\sum_{\substack{i,j \\ i\neq j}} \phi_{(t_i t_j)}(r_{ij}) + \sum_i F_{(t_i)}(\rho_i) + \frac{1}{2}\sum_{i,\alpha}(\mu_i^\alpha)^2 + \frac{1}{2}\sum_{i,\alpha,\beta}(\lambda_i^{\alpha\beta})^2 - \frac{1}{6}\sum_i v_i^2 \qquad (2)$$

where $v_i = \sum_\alpha \lambda_i^{\alpha\alpha}$ and $t_i$ represents the type of atom $i$. In this work, the atom type can either be U or Nb. The embedding energy is a function of electronic density ($\rho_i$) which is supposed to be the linear superposition of the spherical atomic electronic density ($f_{(t_i t_j)}(r_{ij})$), that is

$$\rho_i = \sum_{j \neq i} f_{(t_i t_j)}(r_{ij}) \tag{3}$$

The local dipole vectors and the quadrupole tensors are

$$\mu_i^\alpha = \sum_{j \neq i} u_{(t_i t_j)}(r_{ij}) r_{ij}^\alpha \tag{4}$$

$$\lambda_i^{\alpha\beta} = \sum_{j \neq i} w_{(t_i t_j)}(r_{ij}) r_{ij}^\alpha r_{ij}^\beta \tag{5}$$

where $r_{ij}$ represents the separation distance between atom $i$ and $j$, $r_{ij}^\alpha$ ($\alpha = 1,2,3$) is the $\alpha$ component of the corresponding distance vector. In this work, the U-Nb potential is constructed based on the potential of U and Nb. Thereby, the expressions of $\phi_{(t_i t_j)}(r_{ij})$, $f_{(t_i t_j)}(r_{ij})$, $u_{(t_i t_j)}(r_{ij})$ and $w_{(t_i t_j)}(r_{ij})$ for pure element depend on the element potential adopted in this work. Specially, the ADP potential of U developed by Starikov et al [55] is adopted due to its high fidelity in describing the dynamic responses of U under the shock up to about 95 GPa [26]. The element potential for Nb adopted in this work is an improved version of the one previous developed by us [30], whose testing result is given in *Supplementary Materials*. Now, the remaining unknown functions for the alloy potential are the $\phi_{(t_i t_j)}(r_{ij})$, $f_{(t_i t_j)}(r_{ij})$, $u_{(t_i t_j)}(r_{ij})$ and $w_{(t_i t_j)}(r_{ij})$ for $t_i \neq t_j$. The pairwise interaction function $\phi_{(t_i t_j)}(r_{ij})$ is taken to be the cubic spline functions. For representation brevity, the subscripts of $\phi_{(t_i t_j)}(r_{ij})$ will be omitted below. But we should bear in mind that the parameters in these functions depend on $t_i$ and $t_j$, $t_i \neq t_j$. Let $\{r_I | I = 1, ..., N\}$ to be the preselected node position of the pairwise function $\phi(r)$, and $y_I = \phi(r_I)$, $M_I = \phi''(r_I)$ for short. Then a cubic spline interpolation function for $\phi(r)$ is adopted:

$$\phi(r) = \frac{(r_{I+1} - r)^3}{6h_I} M_I + \frac{(r - r_I)^3}{6h_I} M_{I+1} + \left(y_I - \frac{h_I^2}{6} M_I\right)\frac{r_{I+1} - r}{h_I} + \left(y_{I+1} - \frac{h_I^2}{6} M_{I+1}\right)\frac{r - r_I}{h_I}, \tag{6}$$
$$(r_I \leq r < r_{I+1}, I = 1, 2, ..., N-1)$$

and

$$\phi(r) = \left[\frac{(r_{N+1} - r)^3}{6h_N} M_N + \frac{(r - r_N)^3}{6h_N} M_{N+1} + \left(y_N - \frac{h_N^2}{6} M_N\right)\frac{r_{N+1} - r}{h_N} + \left(y_{N+1} - \frac{h_N^2}{6} M_{N+1}\right)\frac{r - r_N}{h_N}\right]\varsigma(r), \tag{7}$$
$$(r_N \leq r < r_{N+1})$$

where $h_I = r_{I+1} - r_I$ and

$$\zeta(r) = \sum_{i=0}^{5} \lambda_i \left(1 - \frac{r_c - r}{r_c - r_N}\right)^i \tag{8}$$

The cutoff $r_c$ is defined by $r_c = r_N + k_c(r_{N+1} - r_N)$, $k_c \in (0,1)$. Parameters $\{\lambda_i | i = 0,1,...,5\}$ are selected so that $\zeta(r)$ at $r_N$ or $r_{N+1}$ is a unit, and the first and second order derivative of $\zeta(r)$ at $r_N$ or $r_{N+1}$ are zeroes. Specially, formula (6) is used to extrapolate the value of $\phi(r)$ within $r \in (r_{connect}, r_1)$, where $r_{connect} = 1.5$. For $r \in (0, r_{connect})$, the alloy pairwise interaction is extrapolated by

$$\phi_{(t_i t_j)}(r_{ij}) = u_{(t_i)}\phi_{(t_i t_i)}(r_{ij}) + u_{(t_j)}\phi_{(t_j t_j)}(r_{ij}) \tag{9}$$

where $u_{(t_i)}$ and $u_{(t_j)}$ are chosen so that $\phi_{(t_i t_j)}(r_{ij})$ at $r_{ij} = r_{connect}$ are smooth to the first order. For U-Nb alloy, $u_{(t_i t_j)}(r_{ij})$ and $w_{(t_i t_j)}(r_{ij})$ for the case of $t_i \neq t_j$ are taken to be their original form firstly proposed by Mishin et al [54]:

$$u_{(t_i t_j)}(r_{ij}) = \psi\left(\frac{r - r_c^{(t_i t_j)}}{h^{(t_i t_j)}}\right)\left[d_1^{(t_i t_j)} \exp\left(-d_2^{(t_i t_j)} r_{ij}\right) + d_3^{(t_i t_j)}\right] \tag{10}$$

$$w_{(t_i t_j)}(r_{ij}) = \psi\left(\frac{r - r_c^{(t_i t_j)}}{h^{(t_i t_j)}}\right)\left[q_1^{(t_i t_j)} \exp\left(-q_2^{(t_i t_j)} r_{ij}\right) + q_3^{(t_i t_j)}\right] \tag{11}$$

where $r_c^{(t_i t_j)}$ is the cutoff distance for the atom pair between the type of $t_i$ and $t_j$, $d_1^{(t_i t_j)}$, $d_2^{(t_i t_j)}$, $d_3^{(t_i t_j)}$, $q_1^{(t_i t_j)}$, $q_2^{(t_i t_j)}$ and $q_3^{(t_i t_j)}$ are potential parameters to be optimized. The cutoff function in Eq. (10) and (11) are

$$\psi(x) = \frac{x^4}{1 + x^4} \tag{12}$$

The atomic electronic density for alloy is taken to be

$$f_{(t_i t_j)}(r_{ij}) = \varsigma_{(t_i)} f_{(t_i t_i)}(r_{ij}) + \varsigma_{(t_j)} f_{(t_j t_j)}(r_{ij}) \tag{13}$$

where $\varsigma_{(t_i)}$ and $\varsigma_{(t_j)}$ are potential parameters.

For the potential of Nb, the same cubic spline interpolation function is adopted for the pairwise interaction (See Eq. (6) and (7)). However, when $r < r_{connect}$, $\phi(r)$ takes the Morse function:

$$\phi(r) = D_0 \{\exp[-2\alpha(r-r_e)] - 2\exp[-\alpha(r-r_e)]\} \tag{14}$$

where $D_0$, $\alpha$ and $r_e$ are potential parameters determined by the continuity condition at $r = r_{connect}$ up to the second order. And formula (6) is extrapolated to obtain the pairwise interaction within the range of $r_{connect} < r < r_N$. Without ambiguity, the subscript of the potential functions will be omitted hereafter in this part. We take $u(r_{ij})$ and $w(r_{ij})$ to be zero. Consequently, the ADP potential reduces to an EAM potential. The embedding energy is expressed as [56]

$$F(\rho_i) = \begin{cases} F_0^* \left( p\sqrt[p]{\rho_i/\rho_e} - \rho_i/\rho_e \right), & (\rho_i < \rho_e) \\ -F_0 \left[ 1 - n\ln(\rho_i/\rho_e) \right](\rho_i/\rho_e)^n, & (\rho_i \geq \rho_e) \end{cases} \tag{15}$$

where $n = 1/p$, $F_0^*(1-p) = F_0$, $p$ and $F_0^*$ are potential parameters. The atomic electronic density for pure element is taken to be

$$f(r_{ij}) = f_0 \left(\frac{r_{1e}}{r_{ij}}\right)^\theta \left(\frac{r_{ce} - r_{ij}}{r_{ce} - r_{1e}}\right)^2 \tag{16}$$

where $f_0$ and $r_{1e}$ are potential parameters, $r_{1e}$ is the cutoff distance for $f(r_{ij})$. Note that $f_0$ is actually an alloy parameter since its value has no impact on the properties of pure elements. For binary alloy systems, $f_0$ for each element could be absorbed into the corresponding alloy parameter. However, only the relative value of $f_0$ has the impact on the alloy systems. This means that there are $n$-1 additional free parameters left for $n$-component alloy.

With the above model framework, the development of U-Nb potential is divided into two major steps. Firstly, the potential parameters of the EAM potential for Nb are optimized in order to match several key properties of Nb, that is lattice constant, cohesive energy, single vacancy formation energy, elastic constants, body moduli, surface energy, stacking fault energy, twin formation energy, melting point and the equation of state (EOS) for BCC-Nb. These properties of Nb are collected from literatures, either by experimental measurements or first-principles calculations. The testing results of the potential of Nb are shown in *Supplementary Materials*. Secondly, the resulting potential of Nb, alongside with the ADP potential of U [55], are employed to developed the U-Nb alloy potential. In this stage, only the alloy potential parameters are optimized in order to match the physical properties of U-Nb alloy. In contrast to the case of pure elements, huge number of physical properties could be chosen for alloys. Consequently, which alloy properties need to be chosen is of key importance for developing a faithful U-Nb alloy potential. After several trials, we find that the excess mixture energy and the corresponding pressure rising due to the solution substitution are the key reference properties. In order to investigate the dynamic behaviors of $\alpha''$ U-6Nb alloy, the equation of states up to 100 GPa are also included in the fitting database. The selected alloy properties are obtained from the first-principles calculations performed in this work. All fitting procedures are achieved within the software framework of CMOFP developed by us [56, 57]. Ultimately, the optimized potential parameters for Nb-Nb and U-Nb potential are obtained and listed in TABLE II and TABLE III, respectively. And the corresponding potential functions are shown in FIG. 2.

TABLE II. Potential parameters for Nb-Nb.

| Parameter | Value | Parameter | Value |
| --- | --- | --- | --- |
| $n$ | 0.946330607 | $\theta$ | 2.91850789 |
| $F_0$ | 4.92202094 | $k_c$ | 0.50348 |
| $\phi(2.86048)$ | -0.417521 | $\phi(3.80000)$ | -0.147265 |
| $\phi(2.90000)$ | -0.430464 | $\phi(4.67115)$ | 0.0118946 |
| $\phi(3.00000)$ | -0.461317 | $\phi(5.47741)$ | 0.0189608 |
| $\phi(3.15000)$ | -0.486769 | $\phi(5.72096)$ | 0.0164156 |
| $\phi(3.30300)$ | -0.447485 | $\phi(6.16657)$ | 0.0000000 |
| $D_0$ | 0.446824 | $r_e$ | 3.05712 |
| $\alpha$ | 1.16150 | $r_{connect}$ | 2.83188 |

TABLE III. Potential parameters for U-Nb

| Parameter | Value | Parameter | Value |
|---|---|---|---|
| $u_U$ | 10.8621287 | $u_{Nb}$ | -18.9546531 |
| $\varsigma_U$ | 16.3832268 | $\varsigma_{Nb}$ | 189.583591 |
| $f_0^{(U)}$ | 0.00100906324 | $f_0^{(Nb)}$ | 3.02292742e-05 |
| $d_1$ | 122.851025 | $q_1$ | 1.7698006 |
| $d_2$ | 201.70254 | $q_2$ | 1.27496877 |
| $d_3$ | -0.0241401819 | $q_3$ | 8.98913608e-09 |
| $h$ | 2.56008916 | $r_c$ | 6.90236000 |
| $\phi(1.6)$ | 18.888198 | $\phi(3.2)$ | -0.310584905 |
| $\phi(1.8)$ | 15.1583896 | $\phi(3.5)$ | -0.30982553 |
| $\phi(2.0)$ | 10.6354579 | $\phi(3.6)$ | -0.267700228 |
| $\phi(2.2)$ | 5.61812598 | $\phi(3.8)$ | -0.0751982704 |
| $\phi(2.4)$ | 1.7651971 | $\phi(3.9)$ | 0.143877541 |
| $\phi(2.5)$ | 0.615648583 | $\phi(4.0)$ | 0.273007607 |
| $\phi(2.6)$ | 0.155813074 | $\phi(4.2)$ | 0.0109790239 |
| $\phi(2.7)$ | -0.0378043862 | $\phi(4.5)$ | -0.107482333 |
| $\phi(2.8)$ | 0.0616800959 | $\phi(5.0)$ | -0.0496934106 |
| $\phi(2.9)$ | -0.0204647623 | $\phi(5.5)$ | -0.0121704495 |
| $\phi(3.0)$ | -0.128451805 | $\phi(6.90236)$ | 0.0000000000 |

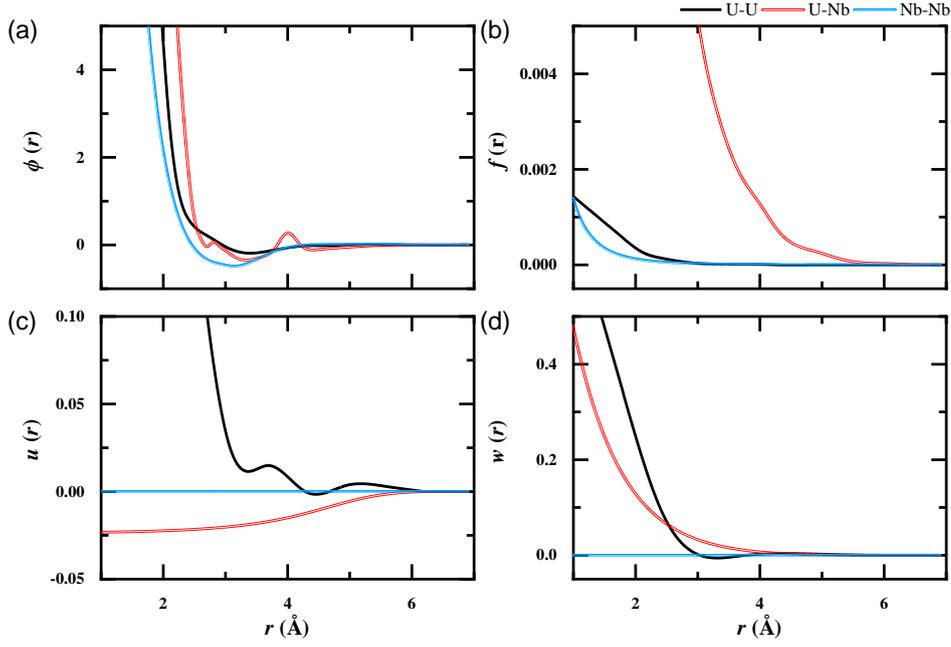

FIG. 2. Potential functions for U-Nb potential: (a) $\phi(r)$ (b) $f(r)$; (c) $u(r)$; (d) $u(r)$.

## C. Testing

As outlined above, we have employed the first-principles calculations to determine the $U_{exc}$ and $P_0$ for various γ U-Nb solid solutions and U-6Nb configurations, with structural parameters detailed in TABLE I. Then, $U_{exc}$ and $P_0$ for these configurations are predicted by the developed ADP potential. The predicted $U_{exc}$ and $P_0$ in comparison with the ones by the first principle calculations are shown in FIG. 3, where the results predicted by the ADP potential (half-filled symbols, dashed lines) exhibit an excellent agreement with the first-principle data (filled symbols, solid lines). For the USF and twin configurations, two doping schemes are employed: (1) uniform random doping across the entire range (0~100 at. % Nb), and (2) spatially localized doping confined to a 6.9 Å width surrounding the stacking fault or twin plane, with a restricted concentration range (0~50 at. % Nb). As shown in FIG. 4, the predicted results by the developed potential also agree well with the first-principle ones.

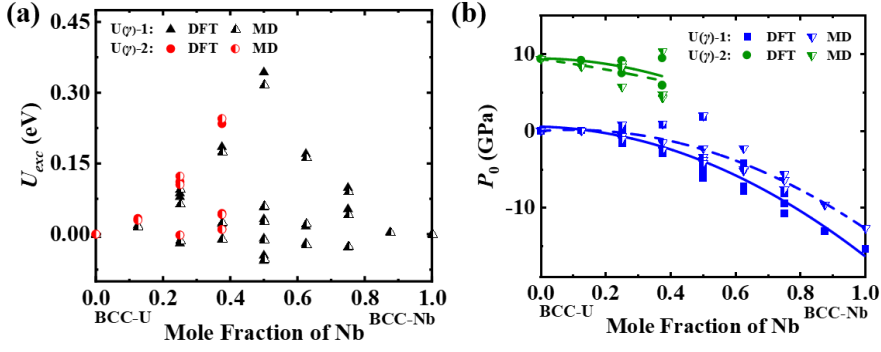

FIG. 3. Comparisons between the predicted results by the ADP and the corresponding first-principle ones. (a) $U_{exc}$ and (b) $P_0$ for U($\gamma$)-1 and U($\gamma$)-2 configurations.

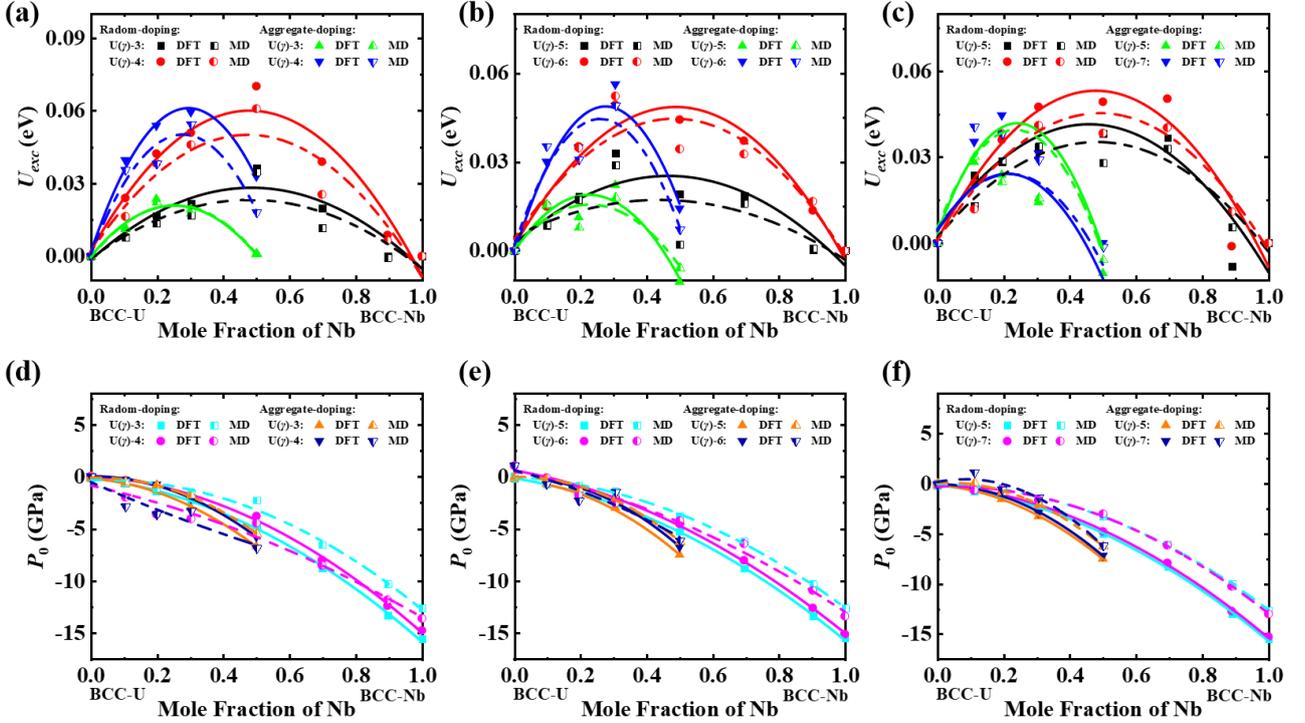

FIG. 4. (a) $U_{exc}$ and (d) $P_0$ for the $\gamma$-U configurations containing {110} <111> USFs, (b) $U_{exc}$ and (e) $P_0$ for the $\gamma$-U configurations containing {112} <111> USFs, (c) $U_{exc}$ and (f) $P_0$ for the $\gamma$-U configurations containing {112} <111> twins and their corresponding perfect configurations as a function of the Nb mole fraction. Solid symbols denote the data from the first principle calculations, while half-filled symbols indicate the results predicted by the U-Nb potential developed in this work. Solid lines are quadratic fits to the first-principle data, and dashed lines are quadratic fits to the predicted data by the potential. The fitted lines are colored identically to their corresponding data points.

Specially, the equation of state and Hugoniot relation of U-6Nb are calculated and compared with experimental results. The equation of state is calculated using an $\alpha$ U-6Nb single crystal with a size of $50a \times 50b \times 50c$ ($a$, $b$, $c$ are the lattice parameters of $\alpha$ phase), totally 500,000 atoms. Periodic boundary conditions and a time step of 1fs are applied to all simulations unless otherwise specified. The U-6Nb single crystal is constructed by randomly replacing 14 at. % of U atoms in $\alpha$-U with Nb and relaxed to zero pressure through an energy minimization via conjugate gradient algorithm. Then the U-6Nb single crystal is further relaxed under NPT ensemble at zero pressure and room temperature for 100 ps. After the relaxation, the U-6Nb single crystal reaches its thermodynamic equilibrium state. Finally, equilibrium volumes under varying pressures are collected and shown in FIG. 5a. The predicted equation of state shows a good agreement with the experimental data [58] up to more than 100 GPa. Besides, Hugoniot relations calculated using large-scale nonequilibirum molecular dynamic simulation method are also examined and compared with experiments (See FIG. 5b). As shown in FIG. 5b, the Hugoniot relation for nanocrystalline U-6Nb samples with an average grain size of ~12 nm, represented by shock wave velocity versus particle velocity ($U_s$-$U_p$), agrees with the experimental one [58] up to a particle velocity of 1.1 km/s, corresponding a pressure of about ~90GPa. Besides, as evident from FIG. 5b, single crystals exhibit strong orientation dependence: loading along $a$-axis shows a classical two-wave structure, loading along $b$-axis lacks elastic precursors, and loading along c-axis displays transient elastic plateaus at low velocities, which are rapidly overdriven by a plastic wave as $U_p$ further increases. The elastic wave propagates is the fastest along $a$-axis due to the minimal interplanar spacing between the adjacent $(100)_\alpha$ planes. Notably, the elastic precursor wave is also not observed in the polycrystalline sample. The absence of the elastic precursor has been widely reported in experiments [13, 15-17], but the underlying microscopic mechanisms remain controversial, primarily due to the lack of direct evidences at atom scale. Recently, the martensitic transformation has been postulated as the dominant mechanism compared to deformation twinning, owing to its associated volumetric collapse that rapidly relieves shear stresses [14, 18, 19]. Our simulation results reveal a displacive α → γ phase transformation mediated by coordinated atomic shuffling along the <100> direction within the (001) basal plane of the parent α-phase. This transformation results in residual {112} <111> deformation twins in the product γ-phase. Detailed analysis of the martensitic phase transition under high pressures is presented in Section IIIC.

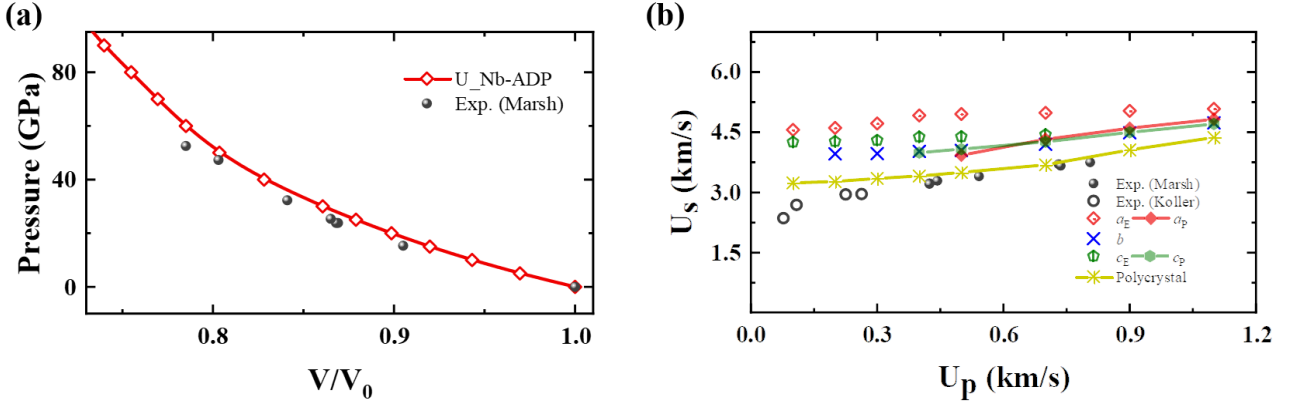

FIG. 5. (a) Equation of state at the room temperature and (b) Hugoniot relation of U-6Nb alloy. Experimental data (from Ref. [14, 58]) are included for comparisons. $V_0$ is the volume per atom at zero pressure. Subscripts $E$ and $P$ represent elastic and plastic waves, respectively.

## III. STRUCTURE PHASE TRANSITION OF U-6Nb ALLOY

### A. α → γ phase transition induced by composition variations

MD simulations are conducted to study α → γ phase transition of U-Nb alloy with varying Nb mole fractions at the ambient condition. First, an initial α-U single crystal with dimensions of $20a \times 20b \times 20c$ is employed, where $a$ = 2.836 Å, $b$ = 5.868 Å and $c$ = 4.935 Å. To construct U-Nb alloys with Nb mole fractions ranging from 0.0 to 1.0, Then a specified fraction of U atoms in the crystal are randomly substituted with Nb atoms. The resulting alloys are relaxed under zero pressure at 300 K for 100 ps. After relaxation, the temperature, pressure, and energy of the alloys stabilize, confirming that equilibrium is achieved. To estimate the miscibility gap in the U-Nb phase diagram, equilibrium states starting from the γ phase with varying Nb mole fractions are similarly generated. For each Nb mole fraction, the equilibrium atomic volume is calculated for systems initialized from both α and γ phases.

The results in FIG.6 indicate that the miscibility gap between the α and γ phases spans a lower limit of approximately 0.04~0.1 and an upper limit of 0.8~0.85, consistent with experimental values (~0.01, 0.84) reported in Ref. [59]. The composition-driven structural phase transition between α and γ phases is directly observable via radial distribution functions (RDFs) and angular distribution functions (ADFs). Chemical bonds within a cutoff radius of 4.3 Å are analyzed. As shown in FIG. 7, the RDF and ADF profiles transition from α-like to γ-like characteristics with increasing Nb mole fraction when starting from the α phase, while the reverse transition occurs when initiating from the γ phase. This behavior conclusively confirms the α ↔ γ phase transformation.

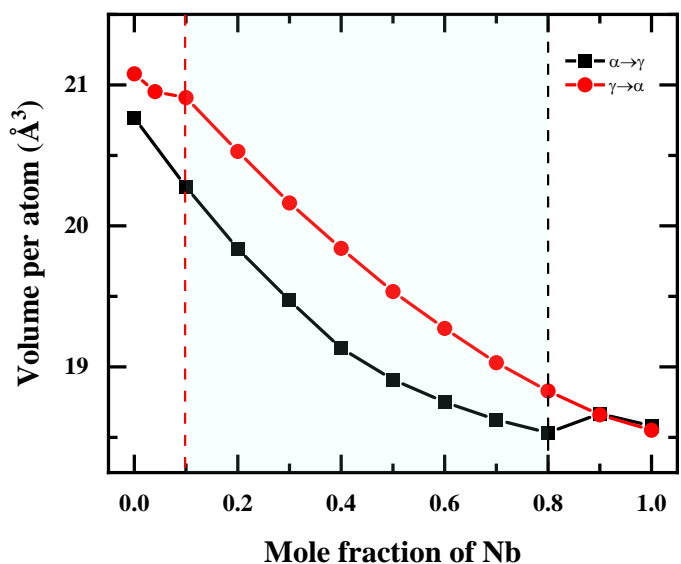

FIG. 6. Volume per atom versus mole fraction of Nb ( $c_{Nb}$ ) under different temperature. The black dash-dotted line is the calculated phase boundary between α and γ phase. The blue shaded region is the miscibility-gap region between α and γ phase.

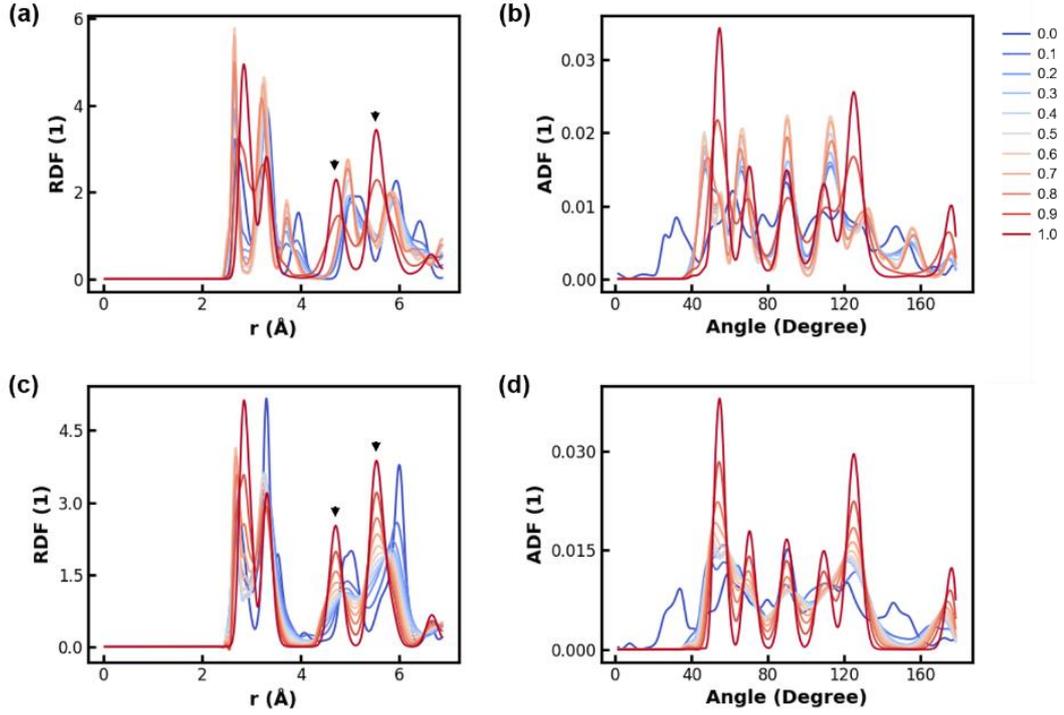

FIG. 7. (a) Radial distribution function and (b) angular distribution function of an initial α-like U-Nb solution alloy with different Nb mole fractions at 300 K. While (c) and (d) are the radial distribution function and angular distribution function for U-Nb alloy starting from γ phase. The black triangles in the figures mark out the characteristic peaks specific to the γ-like lattice.

To make clear the phase transition mechanism of α → γ, the snapshot of α U-90 at. % Nb before and after the phase transition is shown in FIG. 8a. Visualization is achieved in OVITO [60] using its Polyhedral Template Matching (PTM) [61] analysis method with an Root-Mean-Square Deviation (RMSD) of 0.12 (other phases like HCP or FCC account for less than 1%). This method and parameter will be used throughout the rest of this paper. It can be found that γ-phase deformation twins would form after the phase transition. The deformation twins are confirmed as $\{112\}_\gamma$ twins and the orientation relationship (OR) between α and γ phase is identified to be $(001)_\alpha \parallel \{110\}_\gamma$ & $(010)_\alpha \parallel \{112\}_\gamma$. Similar OR is also observed during a rapid cooling of α" U-6Nb alloy by a MTP based MD simulations [35]. It should be noted that the α" → BCT/BCC phase transition of U-6Nb alloy is driven by the temperature, while the α → γ phase transition discussed in this part is caused by the composition variations. FIG. 8b-c detail a two-step phase transition pathway. First, the α phase transforms into an intermediate configuration through $[100]_\alpha$ or $[\bar{1}00]_\alpha$ shuffling among adjacent $(001)_\alpha$ planes. Subsequently, the γ phase forms via shearing on the $(010)_\alpha$ plane along $[100]_\alpha$ & $[\bar{1}00]_\alpha$, accompanied by minor atomic rearrangements. These are consistent with the basic slip system (010)[100] of α-U [61]. Notably, FIG. 7(c) exclusively illustrates the $[100]_\alpha$ shuffling mechanism during the first stage. When shuffling occurs along $[\bar{1}00]_\alpha$, a new variant establishing a mirror symmetry relationship with the $\{112\}_\gamma$ plane of the original phase transition product emerges, leading to the retention of $\{112\}_\gamma$ twins within the post-transformed microstructure. This transformation mechanism induces expansion along the X-direction (*a*-axis) and contraction along the Y-direction (*b*-axis), while maintaining dimensional stability along the Z-direction (*c*-axis), as demonstrated in FIG. 8d. FIG. 8e presents the temporal evolution of the mole fraction of γ phase. Furthermore, FIG. 8f reveals that the volume change associated with the phase transition is nearly zero after ~100 ps, indicating the characteristic timescale of the phase transition process.

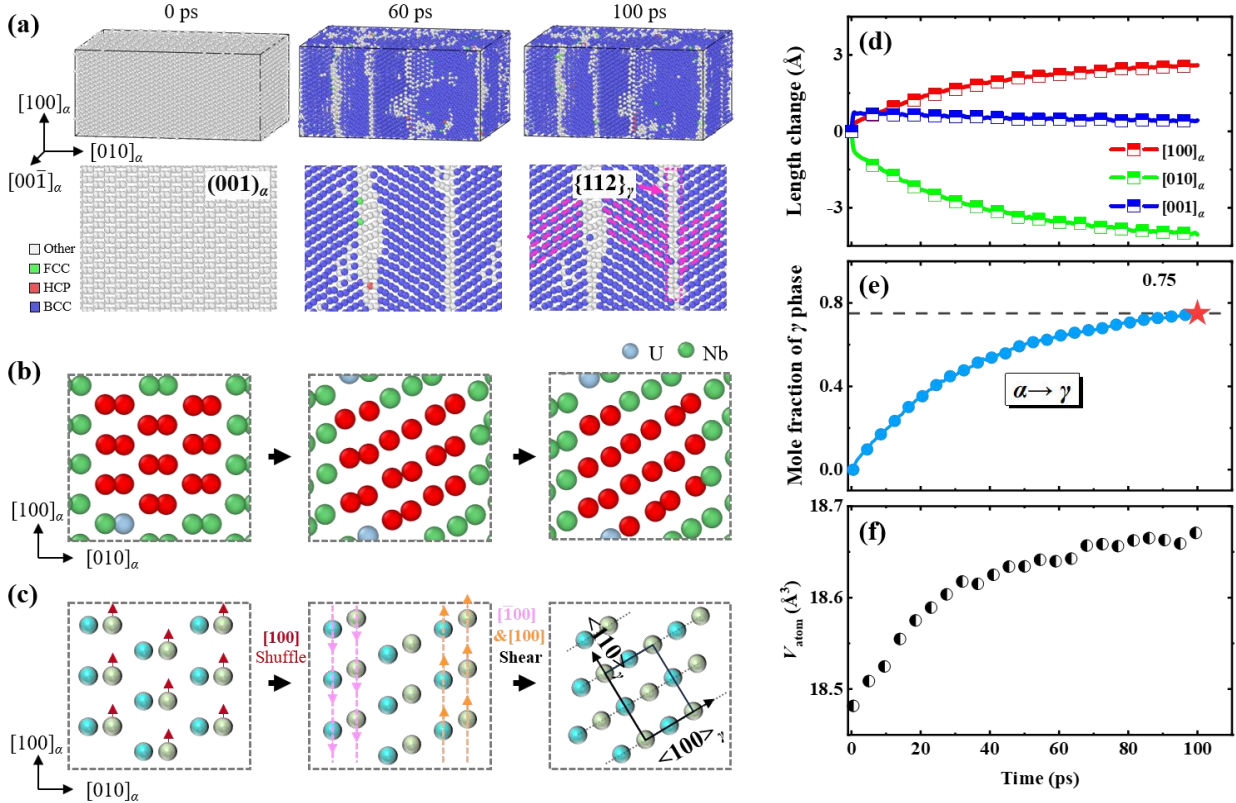

FIG. 8. The α → γ phase transformation in U-90 at. % Nb after relaxation for 100 ps at 300 K. (a) Snapshot of the alloy system and the corresponding $(001)_\alpha$ view at 0, 60 and 100 ps. The color legend is provided on the left. The formed twins are typical $\{112\}_\gamma$ twins. (b) Atomic snapshots extracted directly from the simulation. Blue atoms represent U, and green atoms represent Nb. Specially, the atoms highlighted by the red tracks the process of atom arrangements during the transformation from α to γ phase. (c) Schematics for the phase transition mechanism, where cyan and light green atoms locate on adjacent $(001)_\alpha$ planes. Temporal evolution of (d) the length changes of the system along $[100]_\alpha$, $[010]_\alpha$ and $[001]_\alpha$, (e) mole fraction of γ phase and (f) atomic volume ($V_{atom}$). The red star marks the final mole fraction of γ phase.

## B. Phase transition of U-6Nb at elevated temperatures

Melting points of U-Nb solution alloys with different Nb mole fractions are calculated through MD simulations. From the experimental diagram of U-Nb alloy [59], U-Nb alloys stabilize in γ phase before melting. To examine the stable phase of U-Nb alloy before melting, U-6Nb is taken as an example. An α″ U-6Nb random alloy, having a size of $20a \times 20b \times 20c$, is created and fully relaxed under zero pressure and 300 K. Then the alloy is gradually heating up to 2100 K within 2100 ps with a heating rate of about 0.857 K/ps. Periodic boundary condition is applied during the whole simulations. Average volume per atom versus temperature during the heating is shown in FIG. 9. Five distinctly different volume expansion coefficients (or the slope of the atomic volume versus temperature) could be identified from the result. They are: $\alpha''$ (300 ~ 700 K), intermediate state (700 ~ 1200 K), $\gamma$ (1200 K ~ 1675 K), $\gamma$ + liquid (1675 ~ 1760 K) and liquid (> 1760 K). This agrees well with the experimental phase transition points of ~900K and 1500K~1600K [59]. The structure at 1200 K is shown in the inset of FIG. 9, showing that the phase transition yields a $\gamma$-phase containing twins. Here's phase transformation mechanism is consistent with the composition-induced phase transformation mechanism revealed in Part III A (see FIG. 8c). Further, the RDF and ADF of U-U and Nb-Nb bonds din the U-6Nb alloy within a cutoff distance of 4.3 Å during the whole heating process are calculated. Results shown in FIG. 10 indicates that three different distributions would appear during the heating, which confirms the structural phase transition.

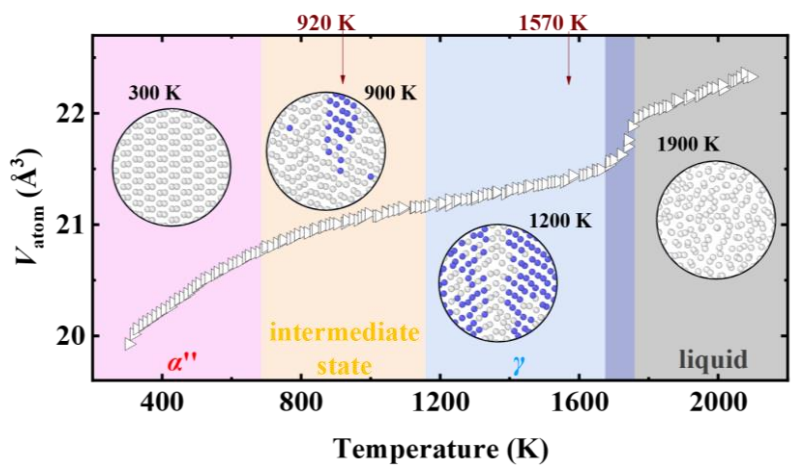

FIG. 9. Temperature-dependent mean atomic volume (gray triangles) for the U-6Nb alloy during heating. The heating rate is 0.857 K/ps. The inset presents atomic configurations at selected temperatures, with $\gamma$-phase atoms colored by blue. The burgundy arrows mark the phase transition temperature reported in the literature [59].

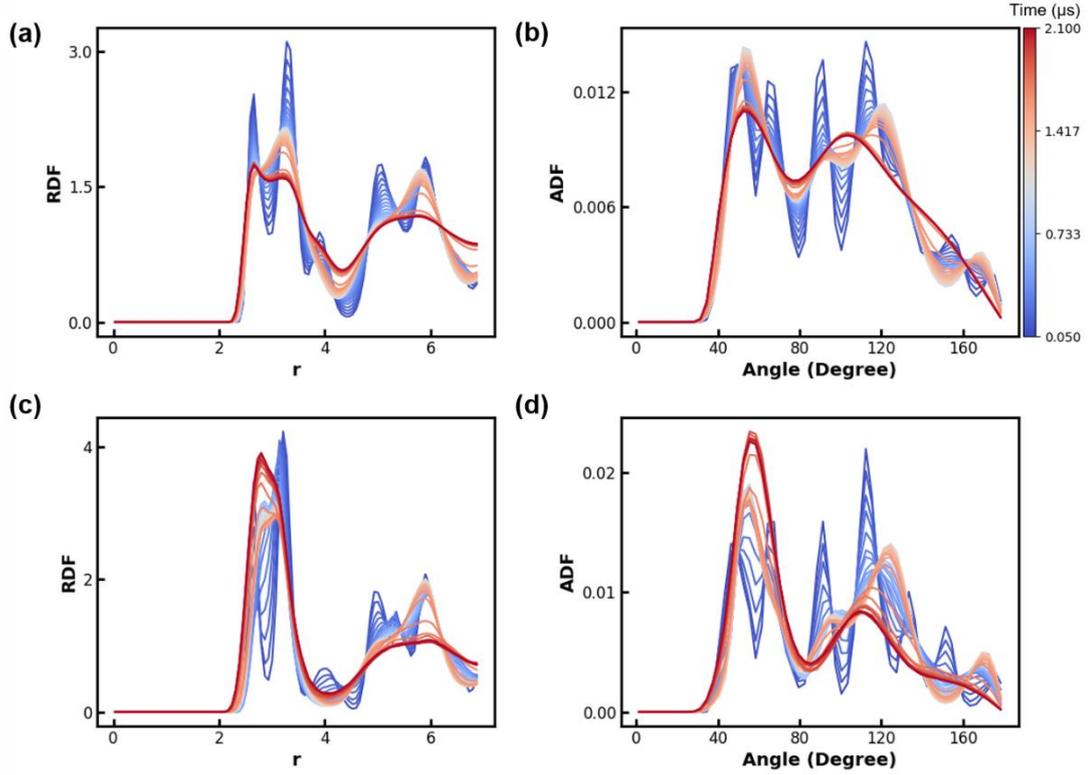

FIG. 10. (a, c) Radial distribution function and (b, d) angular distribution function of different bond types in U-6Nb alloy during the heating simulations, that is, (a, b) for U-U and (c, d) for Nb-Nb bond. It should be noticed that the distribution of U-U and Nb-Nb bonds are not the same in the U-Nb random alloy.

Further, the constant-pressure specific heat capacity of the $\alpha''$ U-6Nb alloy containing 56,000 atoms is calculated as a function of temperature. The alloy is heated from 273 K to 1573 K under NPT ensemble, with a heating rate of ~ 0.5 K/ps. Atomic configurations of the alloy system are recorded every 25 K temperature intervals ($\Delta T$) to enable subsequent derivative calculations of enthalpy at different temperatures. The enthalpy ($H$) for each configuration is calculated through MD simulations for 30 ps, with values averaged over the equilibrium trajectory. The constant-pressure heat capacity ($C_p$) is then determined using $C_p(T) = [H(T+\Delta T) - H(T-\Delta T)]/(2\Delta T)$. The result shown in FIG. 11 exhibits three peaks with the last peak emerging at the position exceeding the temperature range explored (More detailed discussions are presented in Part IIIC). Interestingly, each peak corresponds to a phase region in FIG. 9. Such phenomenon is attributed to the fact that $\alpha'' \rightarrow \gamma$ phase transition takes place in a continuous manner as the volume expansion coefficient is continuous changing during the phase transition (See FIG. 9). The continuous nature of such phase transition is also noticed in a recent work by a machine-learning-based MTP [35].

The specific heat capacity of the $\alpha''$ phase initially rises with increasing temperature. As heating progresses, an increasing fraction of the $\alpha''$ phase transforms into a metastable transient phase, which then evolves into the $\gamma$-phase. This second transformation exhibits slower kinetics than the initial $\alpha''$-to-transient-phase transition, leading to prolonged retention of the transient phase across a broad temperature range (~400–1100 K) under the high heating rates inherent to MD simulations. The endothermic nature of these phase transitions results in an overall suppression of $C_p$, while the peaks observed in FIG. 11 arise from the competition between thermal excitation effects and the two-stage transformation process. Each valley in FIG. 11 approximately corresponds to the onset of a new phase regime: below 700 K is the $\alpha''$ phase dominated region, 700~1100 K is the transient phase dominated region, and 1100 ~ 1600 K is the $\gamma$-phase dominated region. This sequential phase evolution aligns with previous discussions in this part (See FIG. 9).

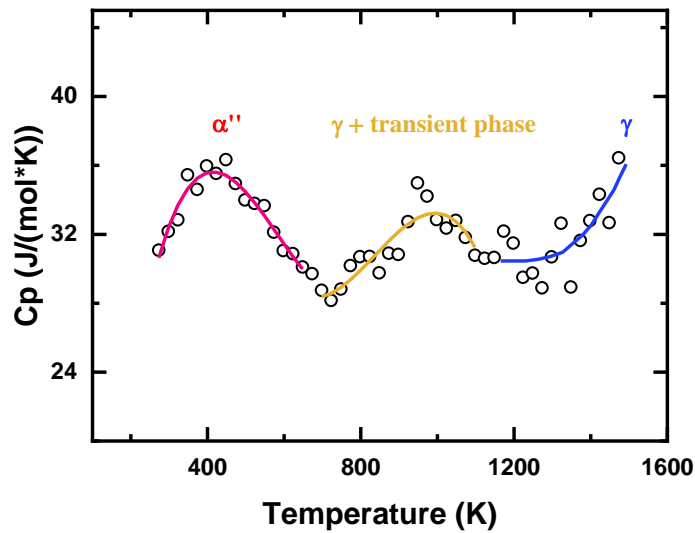

FIG. 11. The calculated constant-pressure specific heat capacity ($C_p$) versus temperature. The colored curves serve as guides for the eye. Particularly, the pink line curve corresponds to the $\alpha''$-phase dominant system, the blue one corresponds to the $\gamma$-phase dominated system and the yellow one represents the system consisting of mixed phases between $\gamma$ and a transient-phase.

Lattice parameters of γ-U-6Nb at elevated temperatures are calculated with compared with the results in literatures. We construct a γ-U-6Nb alloy containing approximately 10,000 atoms. NPT relaxation is performed for 1000 ps at a serial of temperatures between 900 ~ 1100 K under zero pressure. The average lattice parameter over the final 10 ps is calculated for each temperature and shown in FIG. 12 (red symbols), alongside experimental data [62] (black hollow symbols) and machine-learning-based MTP results [35] (tan hollow symbols). The comparison reveals that the lattice parameters of γ-U-6Nb alloy predicted by the ADP potential agree well with the machine-learning-based MTP ones, but slightly smaller than the experimental ones. Furthermore, an accurate linear fit regarding temperature dependence for γ-U-6Nb lattice parameters, based on these data, is shown in the upper right corner of the figure.

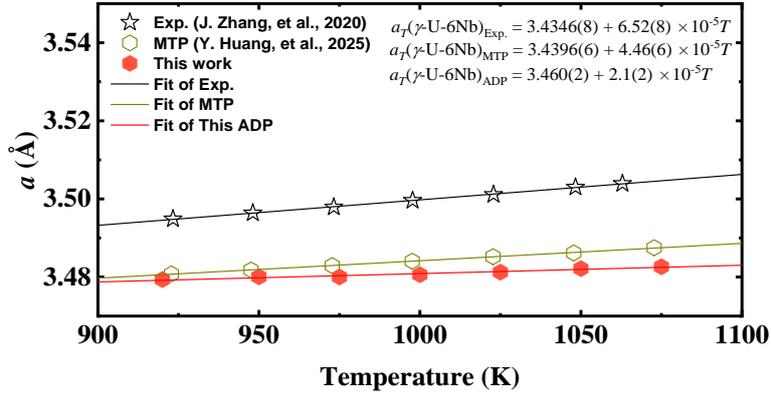

FIG. 12. The lattice parameters of γ-U-6Nb at elevated temperatures are shown, with the ADP-calculated values (red) exhibiting close agreement in both magnitude and trend with experimental data [62] (black) and MTP results [35] (tan).

U-Nb solution alloy initially exists as a γ-phase prior to melting. To estimate its melting points, we evaluate γ-phase stability through a series of fixed-temperature simulations using the solid-liquid coexistence method. Initial alloy structures having a size of $20a \times 20b \times 50c$, with four different Nb mole fractions (0.2, 0.4, 0.6, and 0.8), are constructed. Each sample undergoes relaxation under zero pressure at 1280 K for 100 ps. To establish solid-liquid interfaces, we heat half of each sample along the *c*-direction to 2800 K for 50 ps, followed by a serial of separated relaxations at each trial temperature within [1280, 2800] K for 300 ps. Melting temperatures ($T_m$) are determined by monitoring interface positions through number density profiles along *c* direction, which reveal distinct periodic patterns in solid regions compared to more uniform liquid distributions. The predicted melting points (indicated by red stars in FIG. 13a) demonstrate close alignment with experimental reference bounds shown as black dash-dotted and solid lines. FIG. 13b illustrates representative density profiles at melting points, showing stable solid-liquid interfaces near sample midpoints. For $c_{Nb}$ = 0.8 (FIG. 13c), *γ*-phase atoms (blue) persist in solid regions while liquids contain unstructured atoms (gray). Above $T_m$, the liquid phase encroaches on the solid region, with the reverse occurring below $T_m$. The consistence between the predicted melting points and the experimental data validates the accuracy of the U-Nb potential in modeling the melting behavior of the alloys.

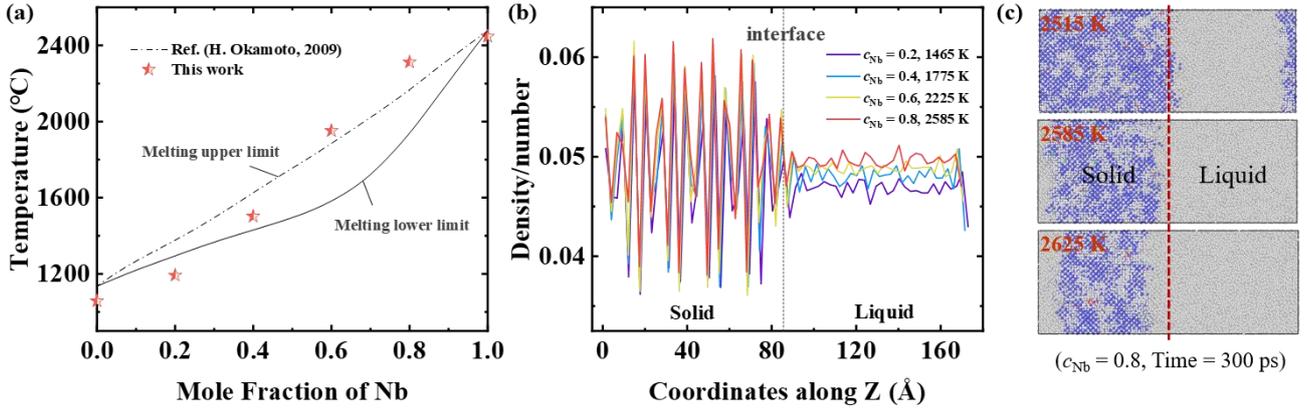

FIG. 13. (a) Predicted melting points (red stars) of U-Nb alloys with varying Nb mole fraction in comparison with the experimental results. (b) Number density profiles along *c* direction after relaxation for 300 ps at the melting points. (c) Atomic snapshots showing the solid-liquid interface migration due to the temperature change for the U-Nb alloy with $c_{Nb}$ = 0.8.

### C. Phase transition of U-6Nb under high pressures

Traditional pressure-induced transformation hypothesis [18, 63] believes that the $α'' \rightarrow γ$-like phase transition may occur at approximately 0.2 GPa. However, recent *in situ* neutron diffraction static high-pressure experiments demonstrate that U-6Nb's monoclinic structure remains stable under static compression up to at least 4.7 GPa at room temperature [19]. This discrepancy highlights U-6Nb's certain compressibility under pure hydrostatic pressure, whereas coupled temperature-pressure space induces complex phase transformation pathways. In shock-wave experiments, the Hugoniot curves of low and high pressures show significant deviations, with the free surface peak stress of the transition range between 5.6 GPa and 10.3 GPa [14]. Moreover, neutron diffraction and TEM data revealed the persistence of the high-temperature $\gamma$ phase in samples post-shock loading [63]. A plausible explanation is the existence of a solid-state critical point for martensitic transformation in the P-T space, where $\alpha''$ might directly revert to the $\gamma$ phase, bypassing conventional reverse martensitic transformation [63]. An experiment had been conducted to verify the existence of this solid-state critical point, finding that the $\alpha'' \rightarrow \gamma^0$ transformation is suppressed under high pressures (1.7 GPa and 2.5 GPa), but no evidence for the existence of a solid-state critical point [19]. The hypothesis currently still lacks direct experimental evidence. In this work, we examine the potential pressure-induced phase transition of $\alpha''$ U-6Nb alloy under adiabatic compressions at atom scale.

As the plastic behavior along the *b*-axis is most pronounced (See FIG. 5b), adiabatic compressions of $\alpha''$ U-6Nb alloy along *b*-axis are conducted. The initial simulation box has a size of $21a \times 10b \times 12c$ and contains around 10,000 atoms, with Nb atoms randomly distributed. Considering the high sensitivity of the adiabatic compressions to pressure, the initial U-6Nb alloy configuration is first subjected to energy minimization and NPT relaxation for 10 ns to balance the pressure and temperature at 0 GPa and 300 K, respectively, with a small enough fluctuation (See *Supplementary Material*). Subsequently, the U-6Nb alloy is relaxed under NVE ensemble for another 5 ns so that the pressure along each principal axis equilibrates around 0GPa (See *Supplementary Material*). Then, compression simulation along its *b*-axis is performed at a strain rate of $1.0 \times 10^8$ s$^{-1}$, yielding a total strain of 0.23 over 2.3 ns. NVE ensemble with three dimensional periodic boundary condition is applied during the simulation. The resulting hydrostatic pressure reached approximately 70 GPa, remaining within the applicable range of the U-Nb ADP potential. To identify the elastic and time-dependent plastic behaviors, the transient states during the compressions are further relaxed under NVE ensemble for an additional 2.5 ns to achieve their equilibrium states. Pressure, temperature, and atomic positions are recorded every 0.25 ps during the last 0.5 ns (yielding 2000 data points per state), with averages serving as the equilibrium values. These equilibrated shear stress (cyan line) and temperature (orange line) versus strain curves, approximately corresponding to the case under low strain rate conditions ($< 4 \times 10^6$ s$^{-1}$), are shown in FIG. 14a. The corresponding transient curve is also included for comparisons. More details could be found in *Supplementary Materials*. Typical time-averaged atomic configurations at point B-E are shown in FIG. 14b, where the $\alpha''$ phase exhibits its characteristic "corrugated" arrangement and a conventional cell of the $\gamma$ phase is highlighted. The result suggests that, during O-B ($\varepsilon < 0.05$), the $\alpha''$ phase essentially undergoes pure elastic deformation with minor lattice distortion and negligible temperature rise, with plasticity initiating at about $\varepsilon = 0.05$ (hydrostatic pressure: $P = 7.9$ GPa, shear stress: $\tau = 1.8$ GPa). Interestingly, there is an elastic relaxation beginning at point A ($\varepsilon = 0.02$, $P = 3.1$ GPa, $\tau = 0.9$ GPa) before the plasticity initiation since the shear stress would slightly decrease after the long time relaxations. Plastic deformation in $\alpha''$ phase begins at B via activation of the (010)[100] slip system and ends at C ($\varepsilon = 0.14$, $P = 30$ GPa). During the plastic deformation, rearrangement of atoms on (001) planes and shear among adjacent (010) planes along [100] or [$\bar{1}$00], with minor atomic spacing adjustments accommodating $\gamma$-phase ordering, ultimately forming a precursor of {112} twins in $\gamma$-phase. Then, during C-D, a progressive transformation from $\alpha''$ phase to $\gamma$ phase occurs, resulting in substantial shear-stress reliefs accompanied by an obvious temperature rising. After D, corresponding to a pressure of ~60.6 GPa, plastic deformation in $\gamma$ phase dominates the deformation process, causing lattice distortions and subsequent hardening of the alloy. However, the fraction of $\gamma$ phase would continue increase at after D when the relaxation time is sufficient long. The corresponding atomic snapshot at O-E as well as the time evolution of the $\gamma$-phase fraction are provided in the *Supplementary Material*. Roughly, point D could serve as an indicator for the completion of the phase transformation.

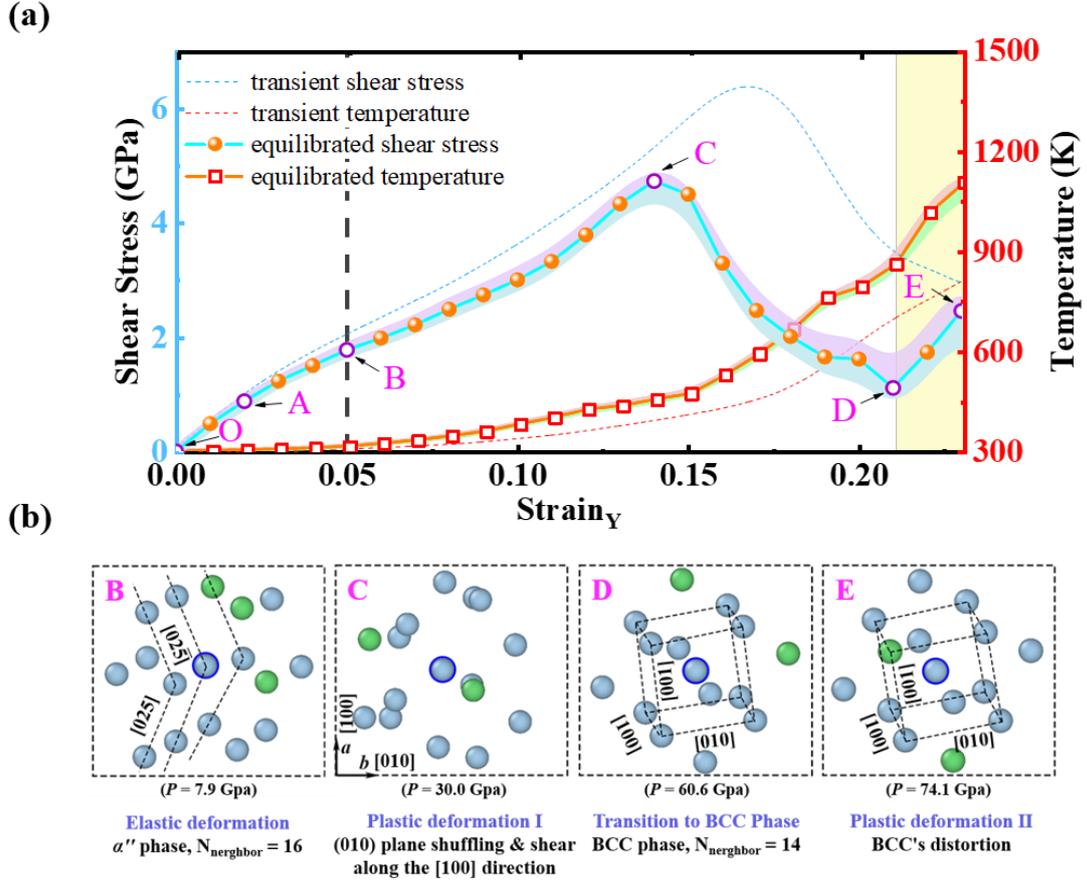

FIG.14. (a) Shear stress and temperature evolution with strain during the adiabatic compression of single-crystal U-6Nb along the *b*-axis. The equilibrium values are obtained by performing additional 2.5 ns relaxations followed by averaging over the last 0.5 ns, with color bands indicating the variation ranges due to the equilibrium fluctuations. (b) Local atomic snapshots after the long time-averaging at points B-E. Only the central atom (blue circle marks), and the neighbors within 4.2 Å are shown in the figures. Blue and green atoms represent U and Nb atoms, respectively.

The RDF and ADF curves of U-U in FIG. 15a-b confirm the existence of BCC phase. Lattice constants for each phase are further determined using a 3D-PCF analysis. The function is defined as $g(\mathbf{r}) = \sum_i \sum_n \delta(\mathbf{r}_n - \mathbf{r}_i - \mathbf{r})$, where *i* represents all atoms within the selected sample range, and *n* represents all neighboring atoms of atom *i* within a cutoff distance of 8.55 Å. Results are shown in the 3D plot on the left side of FIG. 15b. Subsequently, the coordinates of the average atom positions are extracted and shown on the right panel of FIG. 15b. Lattice constants for U-6Nb ($a$ = 2.783 Å, $b$ = 5.782 Å, $c$ = 4.770 Å, y = 0.10) at $\varepsilon_Y = 0$ and $a$ = 3.023 Å for the $\gamma$ phase at $\varepsilon_Y = 0.21$.

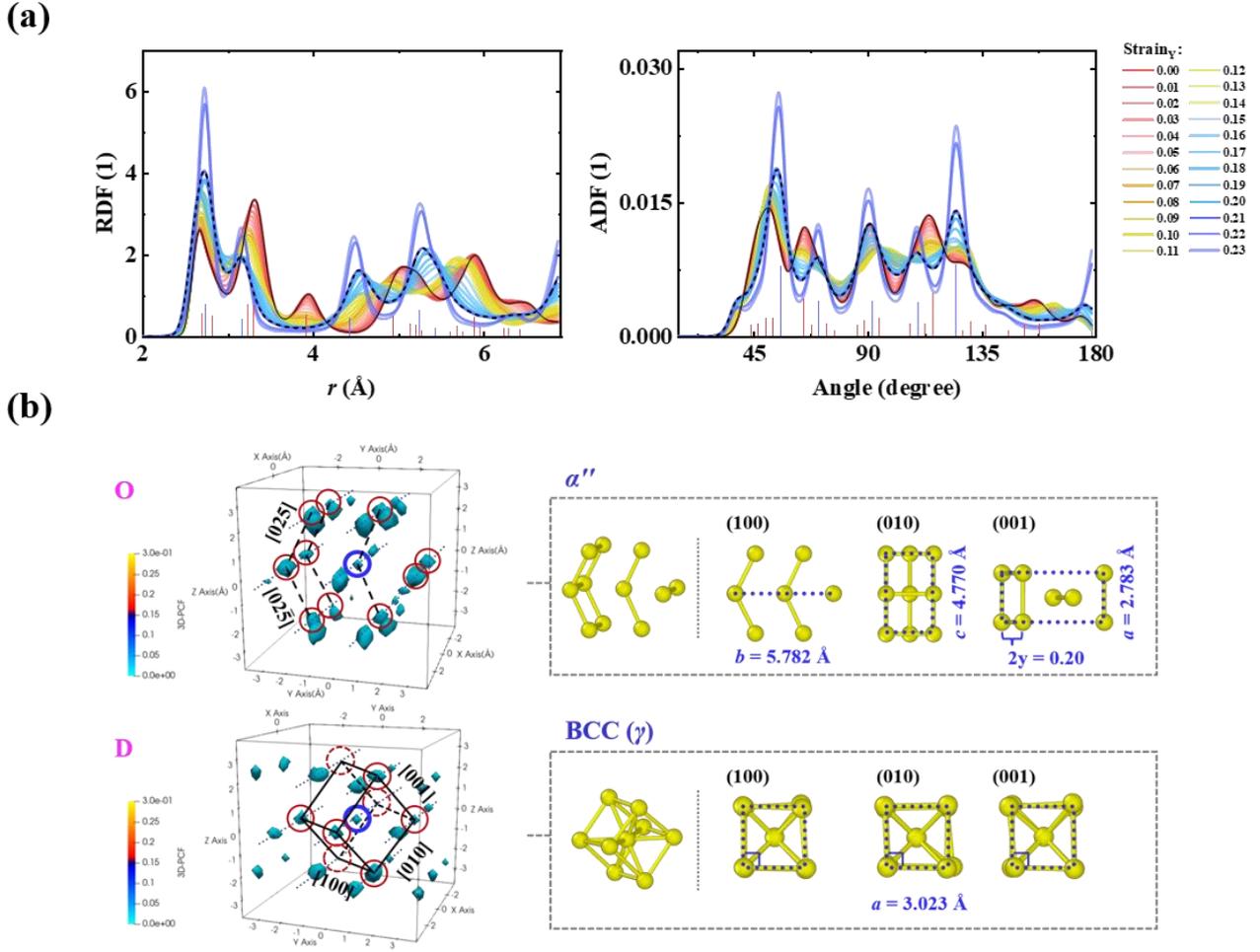

FIG. 15. (a) RDF and ADF curves of U-U about average atomic configurations at equilibrium states under different strains during adiabatic compression of single-crystal U-6Nb along the *b*-axis. Red and blue vertical lines mark peak positions of standard α" and γ phases. Black solid and dashed lines highlight curves at strains of 0 and 0.21 for clearer observation of phase structure evolution. (b) Contour view of the 3D-PCF at points O and D in FIG. 14a. Blue circles indicate central atom, red circles indicate neighbor atoms, and the dashed circles represent the other atoms in a conventional cell but not emerge in the selected region. The selected sample for the γ-phase analyses has a size of 12 Å × 12 Å × 12 Å.

To further confirm the α" → γ phase transition under high pressures, we compares the enthalpy differences per atom between α" and BCC U-6Nb versus pressure at zero temperature through both MD and DFT calculations. The α" U-6Nb sample comprises ~5,000 atoms with Nb randomly distributed and contains 48 atoms with seven U atoms randomly replaced with Nb in the α-U at zero pressure for the MD and DFT calculations, respectively. While the γ U-6Nb sample is derived from the corresponding α″ sample based on the α″ → γ transformation mechanism mentioned above. Single-point energies are then calculated under uniform homogenous compressions. Results shown in FIG. 16 suggest that the ADP potential predicts the threshold pressure of 54.5 GPa, which is slightly lower than that predicted by the DFT calculations (i.e., 67.2 GPa). It should be noticed that the threshold pressure predicted by the ADP potential agrees well with the one observed during the adiabatic compressions (See FIG. 14, point D, whose pressure is 60.6 GPa). According to above discussions, it could be concluded that pressure alone can trigger the α″ → γ transformation in U-6Nb alloy.

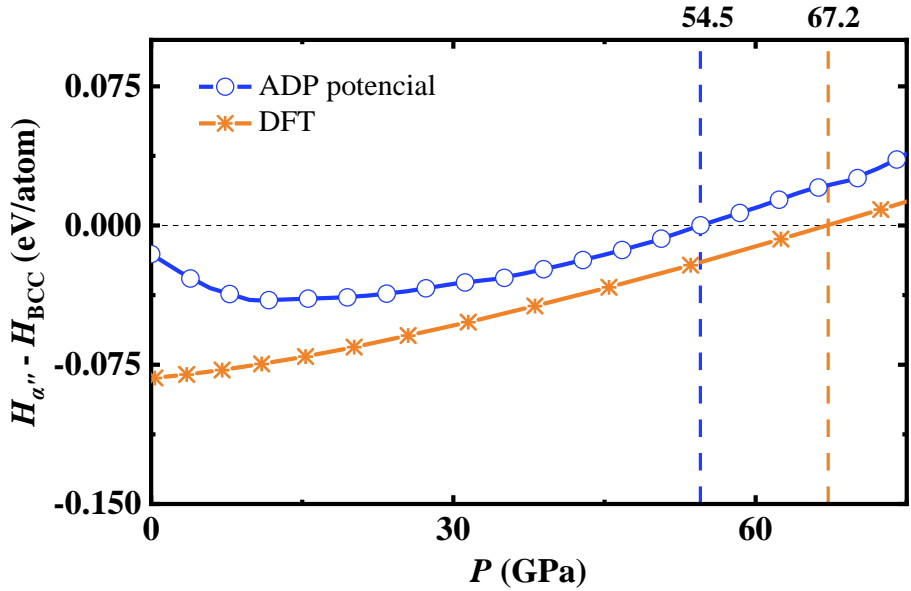

FIG. 16. Enthalpy difference per atom between α″ and γ U-6Nb at 0 K: U-Nb ADP potencial (blue) and DFT (orange). The dashed lines mark the predicted phase transition pressure.

## IV. SUMMARY AND CONCLUSION

In this work, we develop an angular-dependent potential for the U-Nb system based on an existing ADP for U and a new EAM potential for Nb by us. Through fitting of the flexible numerical cross-interaction functions and the remaining alloy parameters to experimental and first-principles data, it could correctly describe α ↔ γ phase transition due to solution concentration variations and the temperature-induced α″ → γ transition in U-6Nb random alloy, except for the basic properties, for example, elastic properties, mixed enthalpy and defect properties (stacking faults, twin and solution-induced pressure). The predicted melting points of U-Nb solid solutions at ambient pressure align with experimental data. Particularly, the equation of states and the Hugoniot relationship of the U-6Nb alloy is correctly predicted up to ~90GPa. The potential also well describes the temperature-dependence lattice parameter of $\gamma$ U-6Nb. And $\alpha'' \rightarrow \gamma$ transition of U-6Nb random alloy observed in static high-pressure experiments is also predicted. Specially, the mechanism for phase transition induced twining in U-6Nb alloys are uncovered. It is found that the $\alpha'' \rightarrow \gamma$ transition couples with the plastic deformation—twinning precursor, ultimately forming characteristic $\{112\}_\gamma$ twins with a molar fraction reaching 75% or higher. Although this complete transformation appears "instantaneous" macroscopically, it appears relatively "prolonged" at the atomic scale, requiring at least nanoseconds to complete. At a strain rate larger than ~$10^6$ s$^{-1}$, a two-stage $\alpha'' \rightarrow \gamma$ phase transition mechanism is uncovered: Firstly, $(001)_{\alpha''}$ shuffling along $[100]_{\alpha''}/[\bar{1}00]_{\alpha''}$ and $(010)_{\alpha''}$ shearing along $[100]_{\alpha''}/[\bar{1}00]_{\alpha''}$ to form an intermediate phase, followed by further relaxation to form the $\gamma$ phase. The orientation relationship is identified to be $(001)_{\alpha''} \| \{110\}_\gamma$ or $(010)_{\alpha''} \| \{112\}_\gamma$. The adiabatic compression simulations suggest that $\alpha''$ sing crystal U-6Nb yields at the pressure of 3.1 GPa, corresponding to the shear stress of 0.9 GPa and transforms into $\gamma$ phase at 60.6 GPa. Additional first principle calculations predict the phase transition pressure to be 67.2GPa, supporting the atomic simulation result. Thereby, it can be concluded that the developed potential could be faithfully applied in U-Nb alloy, especially at the Nb concentration of ~6 wt.%, under a wide range of temperatures up to the melting point and pressures up to ~90GPa. Combined with the atomic simulations, it will significantly promote the fundamental understanding on the mechanical behaviors of U-Nb random alloys, resolve the discrepancy in U-6Nb's pressure-induced phase stability and thus provide valuable reference and theoretical support for the alloy's nuclear engineering applications.

## ACKNOWLEDGMENTS

This work is financially supported by National Defense Science and Technology Key Laboratory

Foundation (Grants Nos. JCKY2022012002) and National Natural Science Foundation of China (Grants Nos. 52471008).